\documentclass[conference]{IEEEtran}
\IEEEoverridecommandlockouts
\usepackage{cite}
\usepackage{amsmath,amssymb,amsfonts}
\usepackage{algorithmic}
\usepackage{graphicx}
\usepackage{textcomp}
\usepackage{xcolor}

\def\BibTeX{{\rm B\kern-.05em{\sc i\kern-.025em b}\kern-.08em
    T\kern-.1667em\lower.7ex\hbox{E}\kern-.125emX}}

\begin{document}

\title{Universal Scanning GUI Tool for Available and Usable TV White Space (TVWS) Spectrum}
\author{\IEEEauthorblockN{Muneer Al-ZuBi and Mohamed-Slim Alouini}
\IEEEauthorblockA{\textit{The Computer, Electrical and Mathematical Science and Engineering Division (CEMSE)} \\
\textit{King Abdullah University of Science and Technology (KAUST)}\\
Thuwal, Saudi Arabia\\
muneer.zubi@kaust.edu.sa; slim.alouini@kaust.edu.sa}
}

\maketitle

\begin{abstract}
In this era of advanced communication technologies, many remote rural and hard-to-reach areas still lack Internet access due to technological, geographical, and economic challenges. The TV white space (TVWS) technology has proven to be effective and feasible in connecting these areas to Internet service in many parts of the world.  The TVWS-based systems operate based on geolocation white space databases (WSDB) to protect the primary systems from harmful interference. Thus, there is a critical need to identify the available and usable channels that can be utilized by secondary white space devices (WSDs) in a specific geographic area.  In this work, we developed a generalized and flexible graphical user interface (GUI) tool to evaluate the availability and usability of the TVWS channels and their noise levels at each geographic location within the analyzed area. The developed tool has many features and capabilities, such as allowing users to scan the TVWS spectrum for any geographic area in the world and any frequency band in the TVWS spectrum. Moreover, it allows the user to apply widely used terrain-based radio propagation models. It provides the flexibility to import the elevation terrain profile of any region with the desired spatial accuracy and resolution. In addition, various system parameters, including those related to regulation rules, can be modified in the tool. This tool exports to an external dataset file the output data of the available and usable TVWS channels and their noise levels, and it also visualizes these data interactively. Moreover, we develop an RF planning and optimization algorithm in this tool to evaluate the various communication metrics between the TVWS BS and the UEs (e.g., WSDs), including received signal strength (RSS), pathloss, signal-to-noise ratio (SNR), data rate, fade margin, and line-of-sight (LOS) availability. This tool also provides an RF optimization technique using Ray Tracing to find the best azimuth and elevation orientation for both the BS and the UE antennas. \end{abstract}


\maketitle

\section{INTRODUCTION}
\begin{figure}[htbp]
\centerline
{\includegraphics[width=\columnwidth]{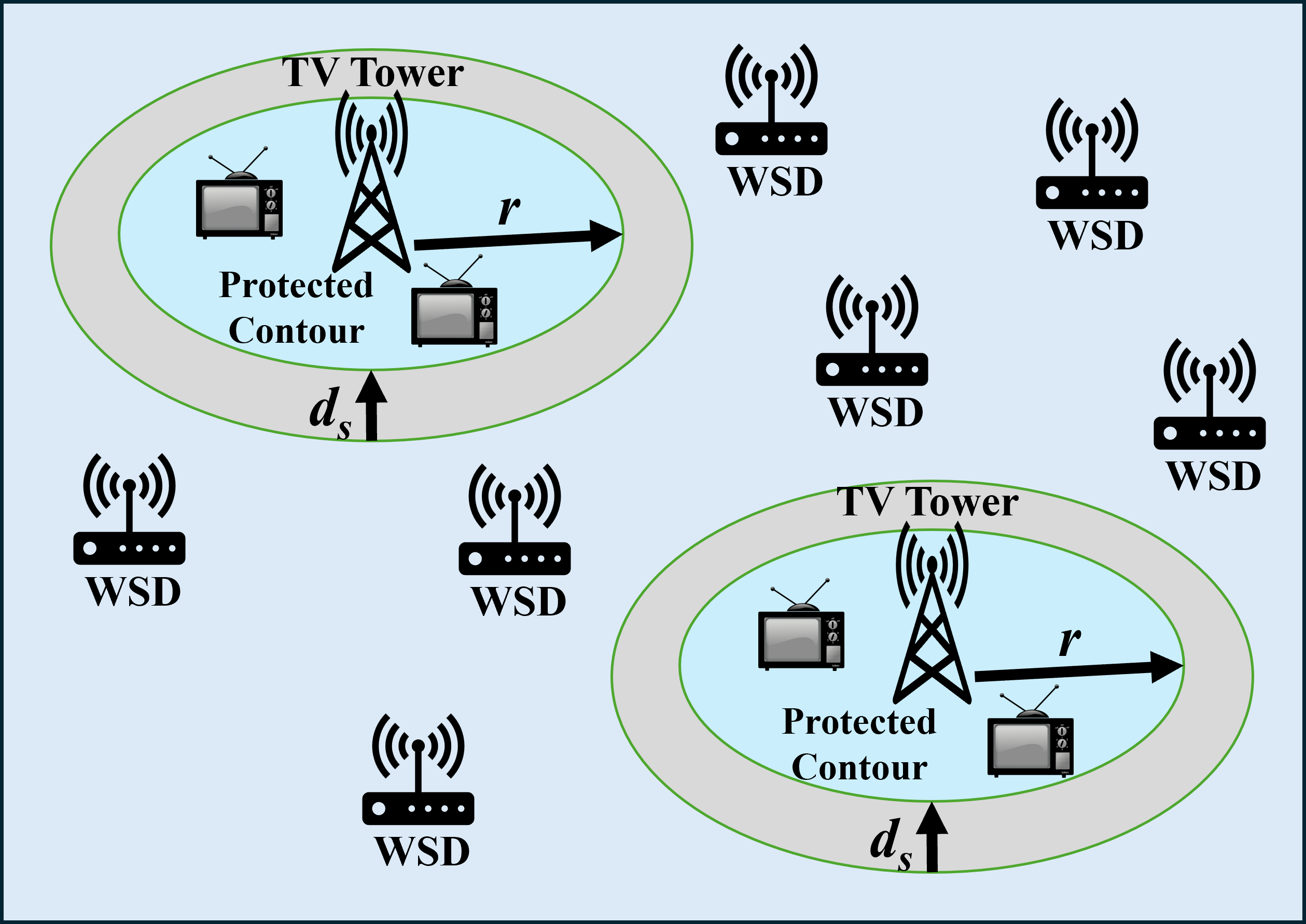}}
\caption{An illustration diagram for the coexistence of the primary TV systems and the secondary TVWS systems.}
\label{fig:fig1}
\end{figure}

Internet connectivity remains a vital necessity in rural and remote areas (e.g., villages, farms, mines, etc.) as accessing the 4G and 5G networks can be a daunting challenge\cite{r1}.  Rural and suburban communities often encounter connectivity hurdles due to challenging terrain, limited financial resources, and a lack of infrastructure. Traditional alternatives such as satellite, microwave, and fibre-optic solutions are cost-prohibitive for these low-income regions, and commercial enterprises naturally seek profitable ventures. Moreover, in an era where wireless communications are becoming increasingly indispensable, efficient use of the radio frequency spectrum has become of utmost importance to solve the spectrum congestion problem. Consequently, TV White Space (TVWS) emerges as an economical solution capable of connecting unconnected rural and hard-to-reach regions and extending 4G/5G coverage to these areas\cite{r17}. Moreover, the TVWS can be used to provide telemetry for the Internet of Things (IoT) applications. The TVWS refers to radio frequency spectrum parts, or white spaces, previously reserved for analogue television broadcasting but have now been available for other wireless communication technologies due to the transition from analogue to digital television technology\cite{r18}. The TVWS spectrum is available in ultra-high frequency (UHF) and very high frequency (VHF) bands with frequencies ranging from 54 MHz to 806 MHz. This spectrum can provide wireless broadband access for secondary (unlicensed) users while protecting the primary users (e.g., digital terrestrial television (DTT), wireless microphone systems, etc.) from interference, as shown in Fig. \ref{fig:fig1}. For example, the regulatory authorities, the Federal Communications Commission (FCC) in the US and the Office of Communications (Ofcom) in the UK, have allowed secondary users to operate in the TVWS band. TVWS technology has various advantages, including low cost, wide coverage area, and energy efficiency. Therefore, TVWS is a promising solution for Internet connectivity in rural and hard-to-reach areas.   

The TVWS system is the first practical application of cognitive radio (CR) technology where the secondary users, aka white space devices (WSDs), share the licensed spectrum with the primary systems, e.g., TV broadcasters and wireless microphones, at a location and time the spectrum not been used by the primary systems. Licensed TV broadcasters usually transmit radio signals over very long distances using transmitters with large power and antenna heights that can highly impact the TVWS-based broadband Internet services.  Moreover, the WSDs may cause harmful interference to the primary systems. Therefore, there is a critical need for a mechanism to test the local area to know the available and usable channels that can be used in a specific area.  The available channel is defined as a vacant channel that is allowed to be used by an unlicensed WSD, while the usable channel is defined as a TV channel with an acceptable noise level. Therefore, not all available channels are usable because some channels may have high noise levels. Different approaches have already been proposed to address this issue, such as spectrum sensing, white space database (WSDB),  beaconing, and hybrid solutions\cite{r19}. Various trials and studies showed the efficiency of the WSDB approach compared to other techniques. The geolocation WSDB  can protect incumbent licensed systems from interference and provide access to the WSDs to use the TVWS spectrum. In the TVWS systems, the WSDs contact the WSDB over the Internet to access the TVWS spectrum. The WSDs send their information (e.g.,  GPS coordinates, antenna height/gain, device type, etc) to the base station (BS), which then contacts the database to request a list of available channels with the corresponding maximum power limit that can be used at these locations. The database provides valuable information on channel availability and noise levels in service areas. Therefore, the regulators have decided to use the WSDBs to coordinate both the primary users and WSDs to reduce and prevent the risk of interference with the primary systems.

Various experimental studies have been conducted to examine the TVWS spectrum at different regions via spectrum scanning using software-defined radio (SDR) based testbeds\cite{r2,r3,r4,r5,r6,r7}. The spectrum sensing techniques in these studies are limited to the areas where the experiments are conducted. Furthermore, it adds additional complexity and cost to the TVWS system if it is embedded in the WSDs. Therefore, regulators and Internet service providers (ISPs) prefer using the WSDB approach by obtaining the available TVWS channels through theoretical and computational techniques. The basic techniques used to obtain the available channels in WSDBs, according to FCC regulation rules, rely mainly on obtaining the coverage areas of the primary systems. The coverage areas of primary broadcasters can be obtained using either the protected service contours, i.e., the  FCC's F(50,50) propagation curves, or the radio propagation models \cite{r23}. In\cite{r8}, the authors evaluate the coverage areas of the incumbent system and available TVWS channels in Ethiopia using the ITU-R P.1546 recommendation-based radio propagation prediction method.  The number of available TVWS channels is evaluated for both fixed and portable devices in the United States, considering the FCC rules and the Longley-Rice propagation model, aka, Irregular terrain model (ITM)\cite{r9}.  In\cite{r10},  the authors proposed a method for estimating the available TVWS  spectrum using the publicly available coverage maps of digital TV (DTV) transmitters in the UK. A study was conducted to obtain the TVWS availability in the 470-790 MHz UHF band in some European countries by examining the impact of the propagation models on the results, including the ITU-R P.1546 prediction method and Longley-Rice model\cite{r11,r12}. In\cite{r20}, a new algorithm is proposed to predict more realistic primary contours based on the Fresnel diffraction theory and the polynomial-based propagation model. Then, the authors used this algorithm to study the TVWS availability in Japan.  A quantitative assessment of the available TVWS in India is calculated using the protection and pollution viewpoints method as well as the FCC regulation rules with the Hata pathloss model \cite{r13}.   In\cite{r14}, a generic methodology for determining the capacity of TVWS using a geolocation-based approach is presented and applied to a sample area in southeastern Europe. A comprehensive evaluation of available TVWS channels in Ethiopia is calculated using a geolocation database front-end (GLSD), developed using the ITU-R P.1546 radio prediction method\cite{r15}.  In\cite{r16},  the authors investigate two approaches based on the regulation rules of the US FCC and the European Electronic Communications Committee (ECC) for using the TVWS spectrum by secondary users to estimate the available TVWS channels. This work shows that the ECC approach is mainly protection-oriented compared to the FCC approach, which focuses on more extensive re-usage of the TVWS spectrum. There are various challenges in evaluating the available TVWS spectrum since it depends on different parameters and conditions of TV broadcasters, propagation channels, terrain profiles,  regulation rules, etc.  The above-mentioned works are mainly focused on evaluating the available TVWS spectrum under specific environmental and regional conditions with fixed system parameters. Moreover, the propagation models used in TVWS channel calculation should be suitable for VHF and UHF bands and should take into account the terrain irregularity, which may affect the accuracy of the estimated number of available and usable channels and noise levels. For example, the significant terrain variation acts as a physical barrier between the primary systems (i.e., the TV systems and wireless microphones) and the secondary TVWS systems, which improves spectrum usage by the secondary systems and thus increases the number of available channels. However, there is a critical need for a generalized graphical user interface (GUI) tool that can flexibly and efficiently generate the available and usable channel dataset for any geographic region around the world using adjustable system parameters. Therefore,  we have developed a generalized GUI tool for obtaining the available/usable TVWS channels and RF  planning of TVWS systems with the following capabilities and features:

\begin{itemize}
\item This tool allows the user to import the terrain elevation data file for any region or country around the world using any digital terrain model (DEM). Thus, it is flexible to be used for any geographical area, with choosing the desired spatial accuracy and resolution of the terrain elevation profile.  
\item  The user can import the TV broadcast towers dataset in the concerned region to be used in the calculation process of channel availability and usability. Users can visualize the spatial distribution of the TV towers on the map.  
\item    The user can control and modify various system parameters. For example, the user can choose a specific value for the spatial pixel resolution,  the WSD antenna height, the TV receiver height, etc. It also allows the user to choose channel bandwidth (e.g., 6MHz or 8MHz) and the frequency band and channel range in the TVWS spectrum.  The user can also exclude the reserved channels from the calculation (e.g., radio astronomy channels). It provides flexibility in specifying the first and last channels in the VHF and/or UHF bands with the corresponding frequencies. 
\item   This tool provides three options for calculating the coverage areas and the RF received signal strength (RSS) using the well-known Longley-Rice propagation model, the terrain-integrated rough Earth model (TIREM™), and the Ray Tracing model. Also, users can adjust the various parameters of these models. 
\item The tool calculates the available and usable TVWS channels and the corresponding noise levels at each geographical location (pixel) and exports the results to an external CSV dataset file. 
\item This tool provides an interactive visualization of the available and usable TVWS channels and the total number of available channels at each coordinate.  Users can search the available/ usable channels for any location by simply entering its coordinates. 
\item The extracted dataset of the available/usable channels can be easily converted later to an appropriate format to be integrated with cloud-based systems or geospatial map tools. 
\item  This tool provides an RF planning and optimization algorithm to evaluate the various communication metrics between the TVWS BS and the UEs (e.g., WSDs), including received signal strength (RSS), pathloss, signal-to-noise ratio (SNR), data rate, fade margin, and line-of-sight (LOS) availability. This tool also provides an RF optimization technique using Ray Tracing to find the best azimuth and elevation orientation for both the BS and the UE antennas.

\end{itemize}

The rest of the paper is organized as follows. In Section II,  the approach and methodology used to develop this tool are presented. Section III illustrates the various parts and parameters of the GUI tool with an example of how to use the tool. The paper is concluded in Section IV.

\section{METHODOLOGY AND APPROACH}
In this Section, we will demonstrate the developed computation engine in this tool in detail, including the structure, the operation, and the computation algorithm. 

\subsection{TVWS Spectrum Scanning}

\begin{figure}[htbp]
\centerline
{\includegraphics[width=\columnwidth]{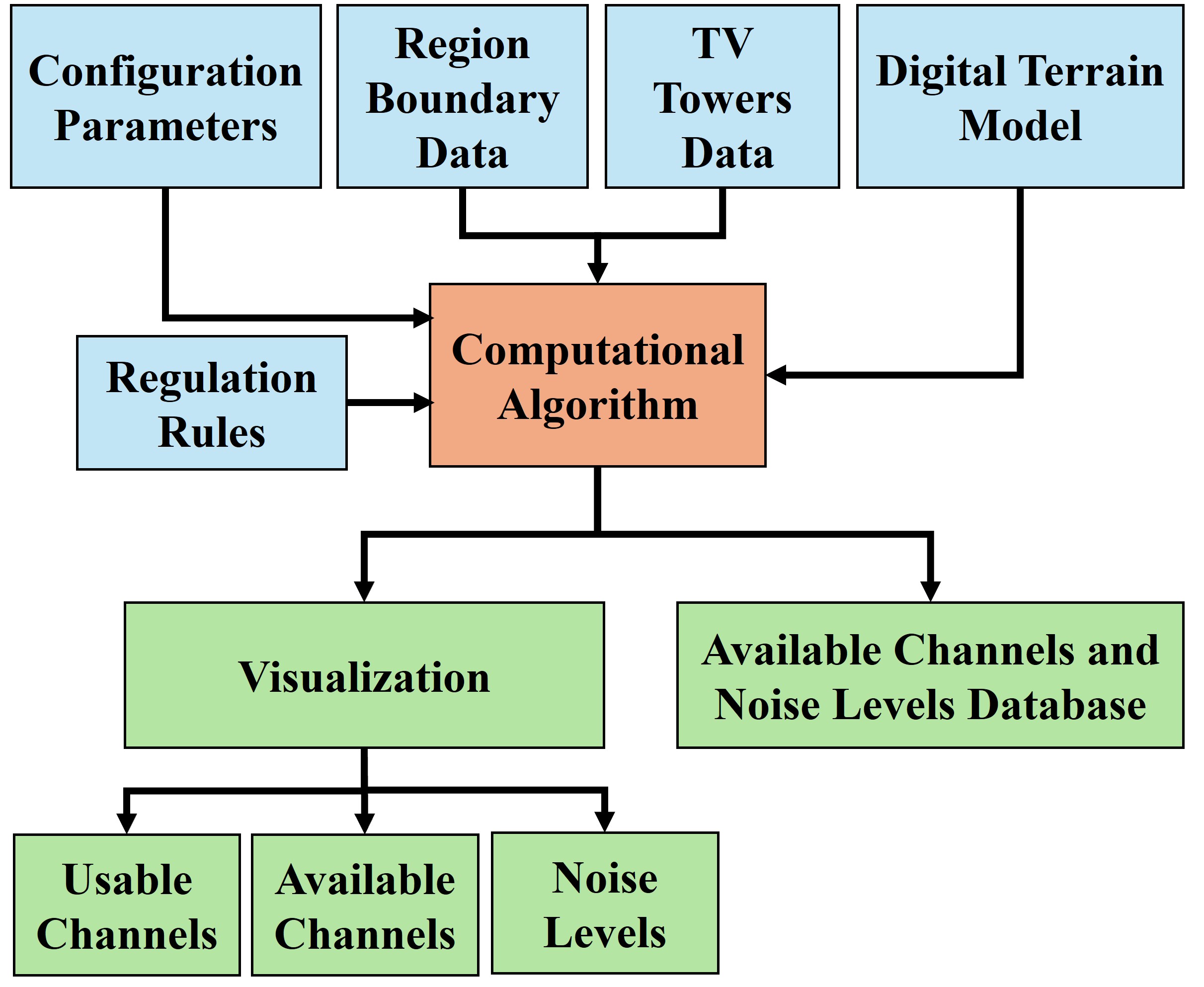}}
\caption{An illustration diagram of the main parts and components in the developed GUI tool.}
\label{fig:fig2}
\end{figure}

An overview of the main components used in this tool to evaluate the TVWS channel availability is shown in Fig. \ref{fig:fig2}. We follow the FCC rules to build this tool, but we make the tool flexible and interactive to fit any scenario and region around the world. In this tool,  the users can adjust the various system parameters.  First, the following data is provided to the computational algorithm: the TV tower data (e.g., coordinates, transmit power, transmit frequencies, antenna heights, emission class, etc.),  the terrain elevation profile, the region boundary, and the radio propagation model. The users can use two terrain-based radio propagation models for calculating pathloss and RSS  with full control of their configuration parameters, and can also import the terrain profile of the concerned region. In addition, the parameters needed in the TVWS systems and the regulation rules are fed into the algorithm such as the reserved channels, the WSD antenna height, the channel bandwidth, the TVWS channels/frequencies under analysis,  the threshold RSS for the TV receivers, the antenna height and gain of the TV receivers,  and the minimum separation distance for co-channels and adjacent channels. The computation algorithm exports the available and usable channels with corresponding noise levels to an external dataset file. In addition, the tool allows the user to visualize this data at any location specified by the user.  

\begin{figure}[htbp]
\centerline
{\includegraphics[width=\columnwidth]{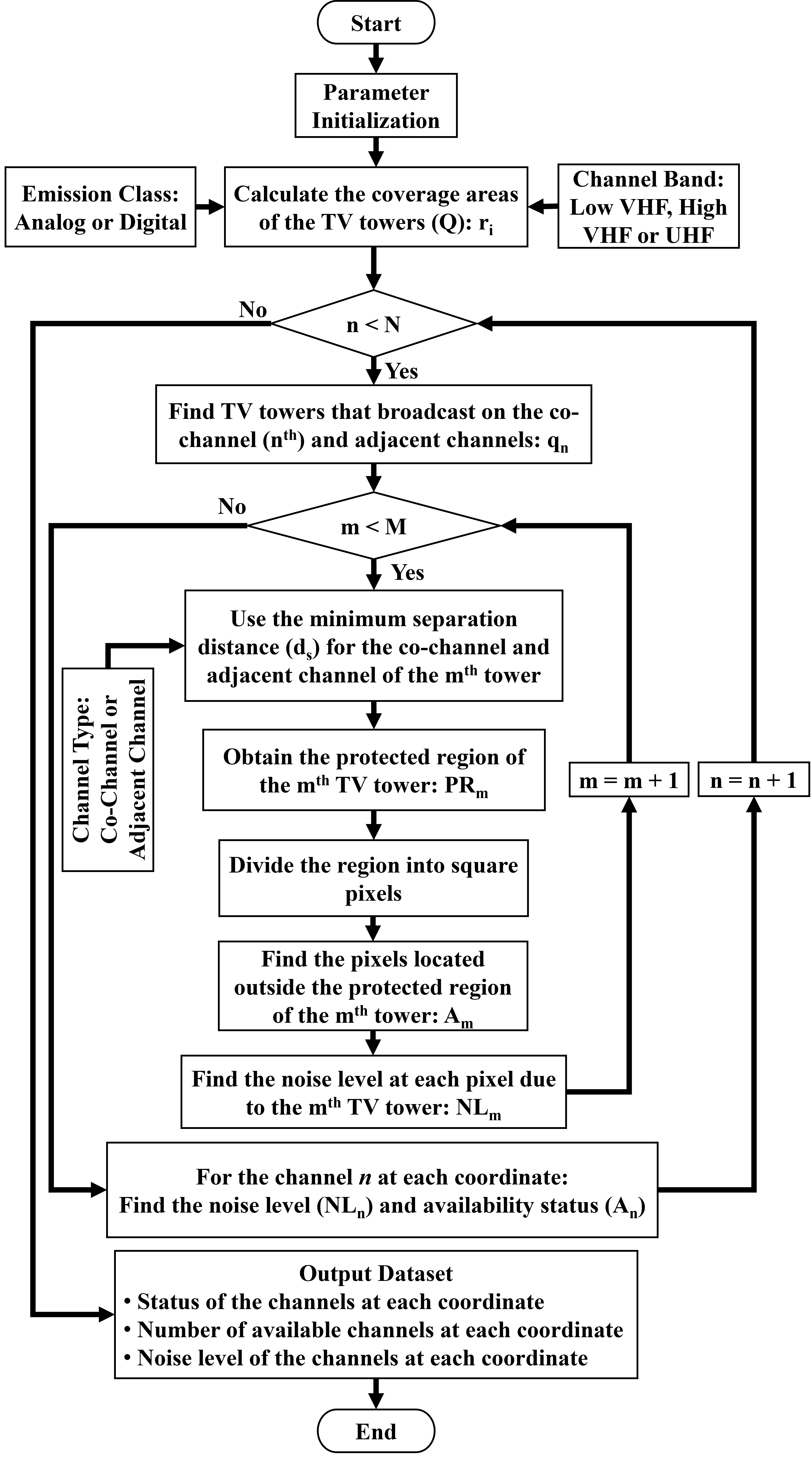}}
\caption{The flowchart of the computation algorithm used in the developed tool.}
\label{fig:fig3}
\end{figure}

Now, we will discuss and demonstrate the developed computational algorithm in detail as shown in Fig. \ref{fig:fig3}. We first obtain the coverage areas (the protected contours) of the TV stations inside the selected region boundary by calculating the RSS using the deterministic terrain-based propagation models, see Fig. \ref{fig:fig1}. These propagation models are highly preferred for obtaining coverage maps and noise levels in wireless communication systems. Here, the user can choose either the Longley-Rice model or the TIREM™ model.  These models were built based on both physics and empirical data and can calculate the pathloss (PL), including free-space loss, diffraction through terrain and obstacles, ground reflection, atmospheric refraction, and tropospheric scatter. It is worth mentioning that the Longley-Rice model is a free and open-source model, while the TIREM™ is a licensed model.  Therefore, to use the TIREM model, the users need to acquire it from Huntington Ingalls Industries to enable access within MATLAB® and consequently in the developed tool\cite{r24}. The RSS at a distance (d) from the i-th TV station can be calculated using the following equations:

\begin{equation} 
RSS_i(d) = EIRP_i +  G_{rx} - PL_i(d) 
\label{eq:1} 
\end{equation}
\begin{equation} 
EIRP_i = P_{tx,i} + G_{tx,i}
\label{eq:2} 
\end{equation}

where $P_{tx,i}$ is the transmit power by the i-th TV station in dBm, $G_{tx,i}$ is the antenna gain of the i-th TV transmitter in dBi, $G_{rx}$ is the TV receiver antenna gain in dBi, and $PL_i(d)$ is the pathloss in dB at a distance $d$ from the i-th TV station, and $EIRP_i$ is the effective isotropic radiated power of the i-th TV station in dBm.  In this tool, the EIRP value is imported from the uploaded TV tower dataset.

 The protected contour of the TV station represents the coverage area in which the TV receivers can detect the minimum signal strength needed for successful audio/video reception. Thus, the coverage area of the TV tower is determined by the farthest distance from the TV station where the received signal power is equal to the threshold RSS ($RSS_{th}$) of the TV receiver. The threshold RSS may have different values depending on the frequency band (e.g., low VHF, high VHF, or UHF) and the emission class of the TV tower (i.e., analog or digital). Therefore, the protected contour radius can be obtained using the following equation:  
 
\begin{equation} 
r_i = max(d) \text{         when      } RSS_i(d) =  RSS_{th} 
\label{eq:3} 
\end{equation}

where $r_{i}$ is the protected contour radius of the i-th TV tower.  
 
For each channel number ($n$) in the channel list to be scanned of size N, the algorithm searches the TV towers that broadcast on the co-channel and the adjacent  channels as follows:
 \begin{equation} 
 q_{n} = Q(ch={n, n\pm1})
\label{eq:4} 
\end{equation}
where $Q$ represents the list of TV stations located within the area being analyzed and $q_{n}=q_{nm}$ for $1\leq m\leq M$ is the list of TV towers that broadcast on the co-channel ($n$) and the adjacent lower/upper channels ($n\pm1$).  

Then, the additional separation distance ($d_s$) is added to the coverage areas of the TV stations in $q_{n}$ list to provide interference protection from the unlicensed WSDs broadcasting on the co-channels and adjacent channels. The resulting overall protected region ($PR_{m}$) for the m-th TV station can be expressed as follows.  
 \begin{equation} 
 PR_{m} = r_{m} + d_s
\label{eq:5} 
\end{equation} 
where $r_m$ is the radius of the protected contour of the m-th TV tower. 
After we reviewed the FCC regulation rules over the past years, we found frequent modifications to the separation distance values.  The separation distance has different values depending on various factors, including the WSD antenna height, the WSD channel (co-channel or adjacent channel), the WSD type (portable or fixed), and the transmit power. Therefore, we have simplified the process by allowing the user to enter the required separation distance for the co-channel and adjacent channel. It is worth mentioning here that the obtained protected contours in this tool, based on the terrain-based propagation models, take into account the effect of the irregular terrain profile, which makes it more accurate than the FCC's F(50,50) propagation curves. The region under analysis is divided into equally square pixels with the corresponding coordinates (latitude, longitude) at their centers. Then, the algorithm finds the coordinates (pixels) that are located outside the protected regions of the m-th tower ($A_m$). 
 \begin{equation} 
 A_{m} = Pixel(Lat, Lon) \not\in PR_{m}
\label{eq:6} 
\end{equation} 
Then, the algorithm calculates the expected noise level ($NL_m$) at various pixels due to the m-th TV tower that broadcasts on the co-channel using Eq.\ref{eq:1}  but with replacing the gain and height of the receiver antenna with those of the WSD instead of the TV receiver. 

The above-mentioned processes are repeated for all TV stations in the list $q_{n}=q_{nm}$ for $1\leq m\leq M$.  Then, the channel is considered available at specific coordinates ($A_n$) if they are located outside the protected regions of all TV towers broadcasting on the co-channel and adjacent channels ($q_{n}$). 

 \begin{equation} 
 A_n = \bigcap_{m=1}^{M} A_{m}
\label{eq:7} 
\end{equation} 
Moreover,  the noise level ($NL_{n}$) of channel $n$ at each coordinate is obtained by finding the maximum noise level among all TV towers that transmit on the co-channel as follows.
 \begin{equation} 
 NL_n = \max_{1\le m \le M} (NL_{m})
\label{eq:8} 
\end{equation} 
Then, the algorithm excludes the reserved channels from the available channel list even if they are found to be available based on the calculation. The reserved channels represent any channel reserved by the regulatory authority for specific usage, such as radio astronomy and wireless microphones. Finally, the obtained data is exported to an external dataset file chosen by the user. The exported data include the coordinates (latitude and longitude) of the pixels inside the region under analysis, with the corresponding total number of available channels, the status of each channel (i.e., available or unavailable), and the noise level. 

\subsection{TVWS RF Planning and Optimization}

In this section, we will illustrate the RF planning and optimization part in the GUI tool. The user can evaluate the various communication metrics between the TVWS BS and the UEs (e.g., WSDs), including RSS, pathloss, signal-to-noise ratio (SNR), data rate, fade margin, and LOS availability. In this tool, the RSS can be calculated using one of the following propagation models: Longley-Rice, TIREM, and Ray Tracing. Then, the path loss ($PL$) is calculated using the following equation for either uplink or downlink. Here, the BS acts as a transmitter and the UE as a receiver in the downlink, while in the uplink, the UE acts as a transmitter and the BS as a receiver. 
\begin{equation} 
 PL = P_{tx} + G_{tx} - L_{tx} + G_{rx} - L_{rx} - RSS
\label{eq:9} 
\end{equation}
where $P_{tx}$ is the transmit power by the BS or the UE in dBm, $G_{tx}$ is the antenna gain of the BS or the UE in dBi, $L_{tx}$ is the cable loss at the BS or the UE in dB, $G_{rx}$ is the antenna gain of the BS or the UE in dBi, $L_{rx}$ is the cable loss at the BS or the UE in dB, and $RSS$ is the received signal strength at the BS or the UE in dBm.  

The SNR is calculated using the following equation in dB:
\begin{equation} 
 SNR = RSS - P_{n}
\label{eq:10} 
\end{equation}
The noise power $P_{n}$ is given in dBm as follows:
\begin{equation} 
 P_{n} = -174 + NF + 10 log_{10}(BW)
\label{eq:11} 
\end{equation}
where NF is the noise figure of the BS or the UE in dB and BW is the channel bandwidth in Hz.

The maximum Shannon throughput (capacity) in Mbps is calculated as follows.  
\begin{equation} 
 C = BW . log_{2}(1 + 10^{SNR/10}) . 10^{-6}
\label{eq:12} 
\end{equation}
Finally, the fade margin (FM) in dBm is calculated using the following equation:
\begin{equation} 
 FM = RSS - RSS_{min}
\label{eq:13} 
\end{equation}
where $RSS_{min}$ is the receiving sensitivity of the BS or the UE in dBm. 

In a point-to-multipoint (PtMP) scenario, the same procedure is followed to calculate these metrics, but for multiple receiver locations. 

This tool also provides an RF optimization technique to find the best azimuth and elevation orientation of the antennas for both the BS and the UE. It uses the Ray Tracing propagation to find the RSS between the BS and UE, and then to find the elevation/azimuth that provides the highest RSS at the receiver as follows. 
\begin{equation} 
 \max_{\{a, e\}}{RSS(a,e)}
\label{eq:14} 
\end{equation}
where $a$ and $e$ represent the azimuth and elevation angles in degrees, which have the following ranges: $a$=[-180, 180] and $e$=[-90, 90]. 

\section{THE GUI TOOL AND INSTRUCTIONS FOR USE}
\begin{figure}[htbp]
\centerline
{\includegraphics[width=\columnwidth]{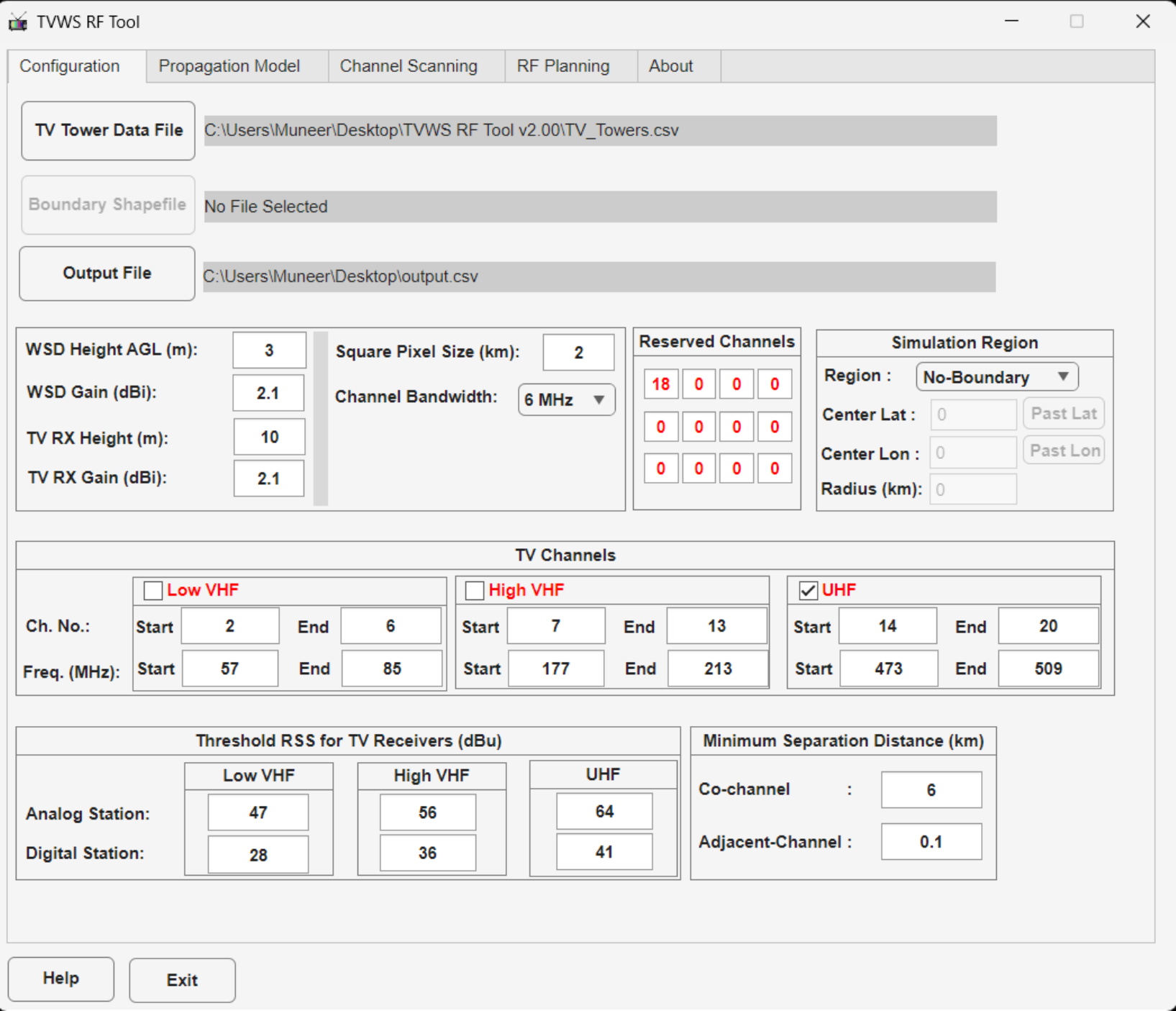}}
\caption{The configuration window in the GUI tool.}
\label{fig:fig4}
\end{figure}

\begin{figure}[htbp]
\centerline
{\includegraphics[width=\columnwidth]{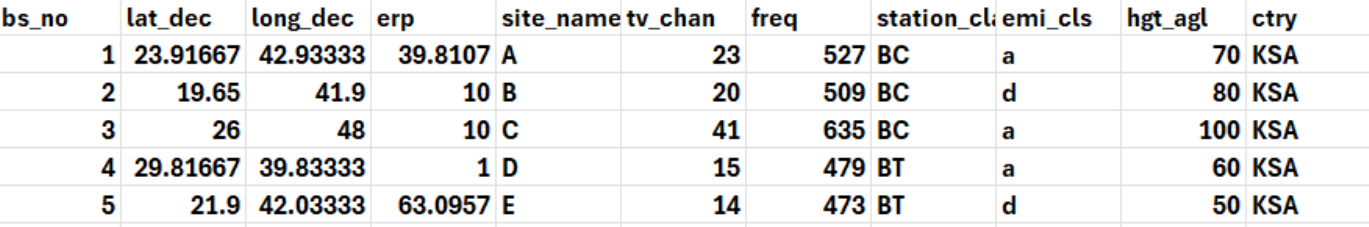}}
\caption{Sample data in TV tower dataset file.}
\label{fig:fig5}
\end{figure}

In this section, we demonstrate the GUI tool we developed, including its windows and components, as well as instructions on how to install and use the tool. This tool is built based on the MATLAB software package, and it can be downloaded from the following GitHub link \cite{r21}.  The user can install the tool by running the downloaded installation file, ensuring Internet access during the installation process in case the associated MATLAB files were not already installed on the computer.  The first window in this tool is the configuration tab, which contains various configuration parameters and files that the user can adjust and choose for TVWS channel scanning, see Fig. \ref{fig:fig4}.   The user should upload a dataset of TV towers in the area under study which includes the following information: TV station index numbers, TV station names, coordinates (latitudes, longitudes) in decimal degrees, effective radiation power (ERP) in kilowatts, channel numbers, center frequencies in MHz, emission class (i.e., analog 'a' or digital 'd'), antenna heights above ground level (AGL) in meters, and the country name. This tool accepts a specific CSV file format of the TV tower dataset, as shown in Fig. \ref{fig:fig5}, which can also be found in the tool installation folder. The user should use the same header title as shown in Fig. \ref{fig:fig5}. The location of the output data file, which includes the status and noise level of the channels, can be selected using the "Output File" button or can be left to the default local folder. The region boundary can be chosen from the "Simulation Region" panel to be one of the following options: No-Boundary, Circle, or Shape-File. The no-boundary option means that the tool will automatically set the region boundary to the maximum distances in the coverage areas of the TV towers. If the Circle option is chosen, the user should enter the coordinates of the center point in decimal degrees and the radius of the circular area in km. However, any arbitrary area shape can be used as a boundary by choosing the shape-file option and uploading the area shape file using the "Boundary Shapefile" button. The user can also set the height and gain of the WSD antenna and the square pixel size in km. As the pixel size decreases, we get a higher spatial resolution, but at the same time, this leads to longer computational time. We think a pixel size in the range of a few kilometres (e.g., 1-3km) is sufficient for such an application where the TV signal may reach distances over 100km, and therefore, usually, there is no significant propagation variation over this pixel size. Another main configuration parameter is the channel bandwidth (6, 7, or 8 MHz) and TVWS channels under analysis, which can be chosen in low VHF, high VHF, or/and UHF bands. The channel bandwidth is chosen based on the bandwidth used in the country where the TV towers are located. The user should select one or more bands via the corresponding check boxes (i.e., low VHF, high VHF, and UHF) depending on the channels that need to be scanned. Then, the start and end values for the channel number and the corresponding center frequency should be entered. Also, the user can enter up to twelve reserved channels to be excluded from the calculation. These reserved channels will be considered unavailable everywhere, but the tool will calculate their noise levels and will include their impact on the adjacent channels. In addition, the tool allows the user to tune the threshold RSS of the TV receiver for analog and digital TV receivers at each frequency band. Users can also adjust the minimum protection separation distance in km for co-channel and adjacent channels.  These parameters can be defined by the regulatory authority in the region or country under analysis.  In addition, this tool allows the user to specify the antenna height and gain of the TV receivers and WSDs.

\begin{figure}[htbp]
\centerline
{\includegraphics[width=\columnwidth]{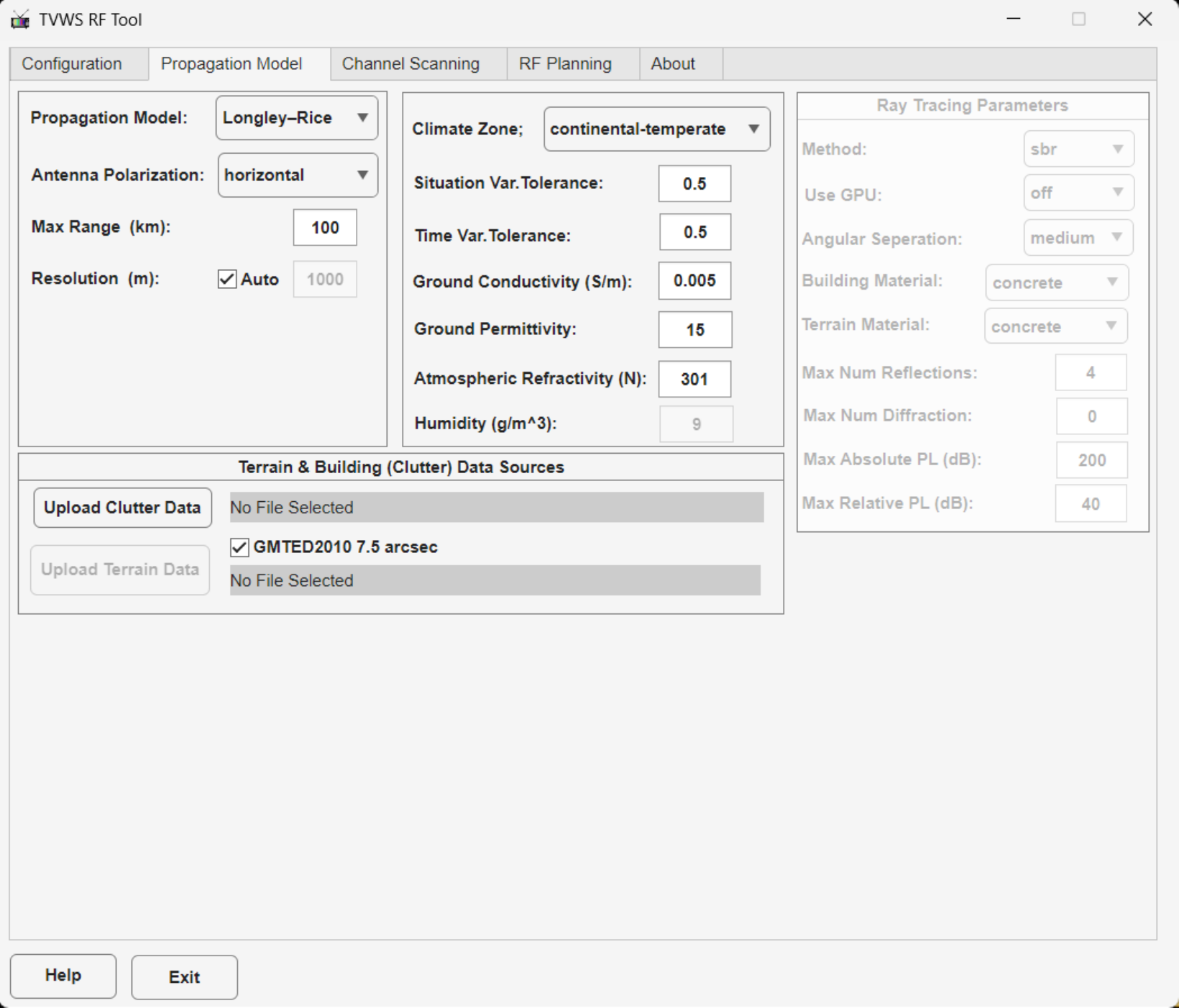}}
\caption{The propagation model window in the GUI tool.}
\label{fig:fig6}
\end{figure}

The parameters associated with the radio propagation models can be adjusted in the propagation model window of the GUI tool, as shown in Fig. \ref{fig:fig6}.   In this window, the user can choose the terrain data source for a specific region to be either imported directly from the Global Multi-resolution Terrain Elevation Data 2010 (GMTED2010) 7.5 arcsec digital elevation model (DEM) or an external digital terrain elevation data (DTED) file of format (.dt0, .dt1, .dt2). For more information on the freely available DEMs and the corresponding online datasets, the reader can refer to this article \cite{r22}.  Here, we will use the default option of GMTED2010 7.5 arcsec.  In addtion, users can upload clutter data for building and obstacles in the area under analysis, i.e., OpenStreetMap (OSM) file, to be considered in the calculation process of RSS and pathloss. The user can choose the Longley-Rice or TIREM model and then adjust the model parameters including the antenna polarization,  situation variability tolerance (0.001 - 0.999),  time variability tolerance (0.001 - 0.999),  ground conductivity in (S/m), ground permittivity, atmospheric refractivity near the ground in (N-Units) (250 - 400),  absolute air humidity near the ground in (g/m$^3$) (0 - 110), and climate zones. The antenna polarization can be either "horizontal" or "vertical" for both transmitter and receiver antennas.  The ground conductivity and permittivity are used by the Longley-Rice model to calculate path loss due to ground reflection and diffraction, and by the TIREM model to calculate path loss due to ground reflection.  The default values shown in the window correspond to the average ground. The models use atmospheric refractivity to calculate the path loss due to refraction through the atmosphere and tropospheric scatter. The default value of atmospheric refractivity corresponds to average atmospheric conditions. The Longley-Rice model uses climate zones to calculate the variability due to changing atmospheric conditions. The user can select one of the following climate zones: continental-temperate, equatorial, continental-subtropical, maritime-subtropical, desert, maritime-over-land, or maritime-over-sea. The default value represents average atmospheric conditions in a particular climate zone. In the Longley-Rice model, the time variability tolerance is the long-term variability of path loss, including atmospheric conditions, while neglecting the short-term variability, such as fast multipath fading. It determines the required system reliability as a fraction of time where the actual path loss is expected to be less than or equal to the model prediction. Also, the situation variability tolerance of the path loss occurs due to uncontrolled or hidden random variables. It determines the required system confidence as a fraction of similar situations where the actual path loss is expected to be less than or equal to the model prediction. Finally, the TIREM model uses the humidity value to calculate path loss due to atmospheric absorption. The default humidity value represents the absolute humidity of air at 15 degrees Celsius and 70\% relative humidity. In addtion to the previously mentioned models, users can also choose the Ray Tracing model for RF planning and optimization.  Then, users can adjust the Ray Tracing parameters, including the method (shooting and bouncing rays (SBR) or image methods), enabling GPU calculation if the PC has a GPU, and angular separation representing the average number of degrees between launched rays. The surface material of geographic buildings and terrain can also be selected.  The users can adjust the maximum number of path reflections and edge diffraction to use ray tracing. The maximum number of path reflections for the image method is 0, 1, or 2, while for the SBR method, it is in the range [0, 100]. The edge diffraction is supported only with the SBR method, and the maximum number of edge diffractions is 0, 1, or 2. The Maximum absolute path loss, in dB, is used to enable the discard of propagation paths that have a path loss larger than an absolute path loss threshold. The maximum relative path loss, in dB, is used to enable the discard of propagation paths that have a path loss larger than the strongest ray by a maximum relative threshold. 

The maximum range parameter determines the maximum radial distance (km) from the TV towers at which the algorithm calculates the received signal strength. This will reduce the computational time, as usually the received signal power is very low after a specific distance from the tower (i.e., the signal can not be detected by the TV receivers). Also, the resolution parameter can be selected as either automatic ("Auto") or a specific distance in meters.  This value specifies the coverage quality by determining the distance between the locations at which the propagation model estimates the received signal strength and path loss.  If the user chooses the "Auto" option, the algorithm computes the maximum value scaled to the maximum range parameter. Otherwise, the model will use the defined distance value, where a smaller value will increase the coverage quality but also will increase the computation time.

\begin{figure}[htbp]
\centerline
{\includegraphics[width=\columnwidth]{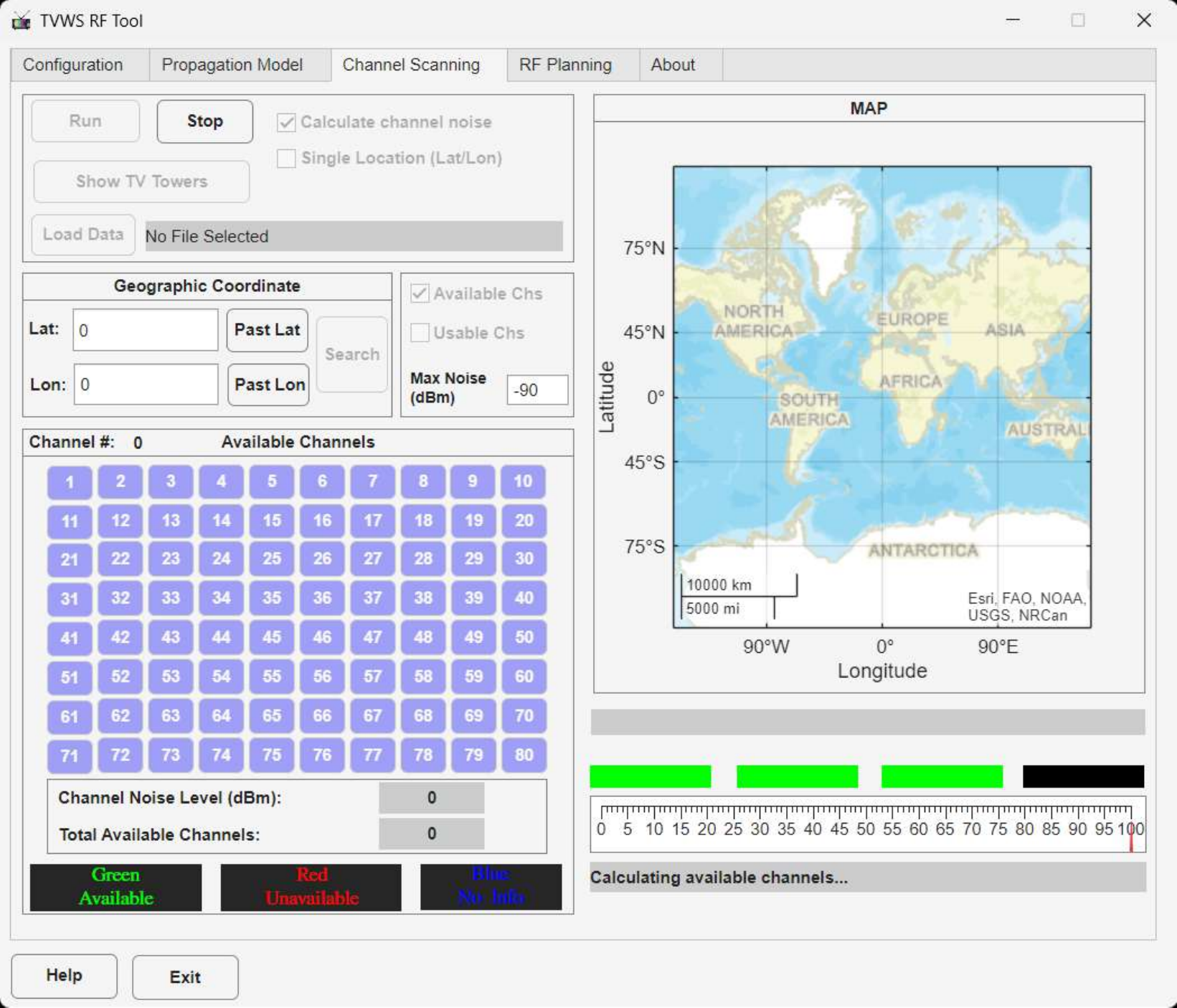}}
\caption{The calculation (simulation) window in the GUI tool.}
\label{fig:fig7}
\end{figure}

\begin{figure}[htbp]
\centerline
{\includegraphics[width=\columnwidth]{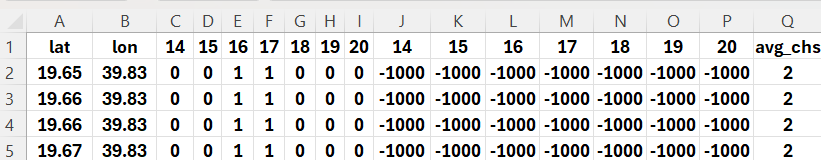}}
\caption{Sample data in the output CSV file after the calculation.}
\label{fig:fig8}
\end{figure}

After the configuration process is completed, we can move to the calculation window to calculate the available/usable TVWS channels and noise levels as shown in Fig. \ref{fig:fig7}. The user has an option ("Calculate channel noise") to include or exclude the noise level calculation from the output results. However, if the noise level calculation is disabled, the output results will only include the availability status of the channels without any information on the noise level and usability of the channels. Thus, we can not decide if the channels are usable or unusable without the noise level calculation. Users can choose either to calculate the channel availability for the whole area or for a specific coordinate using the "Single Location (Lat/Lon)" checkbox. If a single location is selected, the user should enter the coordinates in the designated fields inside the Geographic Coordinate panel. Then, the user can click the Run button to start the calculation (simulation) process. The progress and status bar at the bottom of the window displays information about the calculation progress. The calculation process may take a few minutes to hours, depending on the size of the TV station dataset, the range of channels scanned, and the size of the geographic area being analyzed. For example, it took a few hours to calculate the data for the entire area of Saudi Arabia using a pixel size of 2km for channels 21 to 61 and a dataset of about 700 TV towers. However, in the current example, it took a few minutes because the number of TV towers and scanned channels is small.  At the end of the calculation process, the tool will export the output data to an external CSV file stored in a path predefined by the user during the configuration phase. The output data contains a list of the coordinates and the corresponding status of the channels and the noise levels, as shown in Fig. \ref{fig:fig8}. The first row (header) in this file shows the latitude "lat", the longitude "lon", the channel numbers (e.g., 14 to 20 in this example), and the last column is the total number of available channels at each coordinate. Here, the available channel is represented using the digit "1" and the unavailable channel using the digit "0". The noise level is given in dBm and is set to -1000 if there is no noise in the channel at all. The output data will be loaded directly into the tool. However, the user can upload the output data file at any time later using the “Load Data” button. The tool also allows the user to display the locations of the TV towers on the map using the “Show TV Towers” button.  In addition, the user can select or deselect the options (Available Chs or Usable Chs) to show either the available channel or the usable channels. Also, this tool allows the user to show the channels that are both available and usable by enabling both options. The value of the "Max Noise" parameter can be adjusted to be used as a threshold to decide whether the channel is usable or not. 

\begin{figure}[htbp]
\centerline
{\includegraphics[width=\columnwidth]{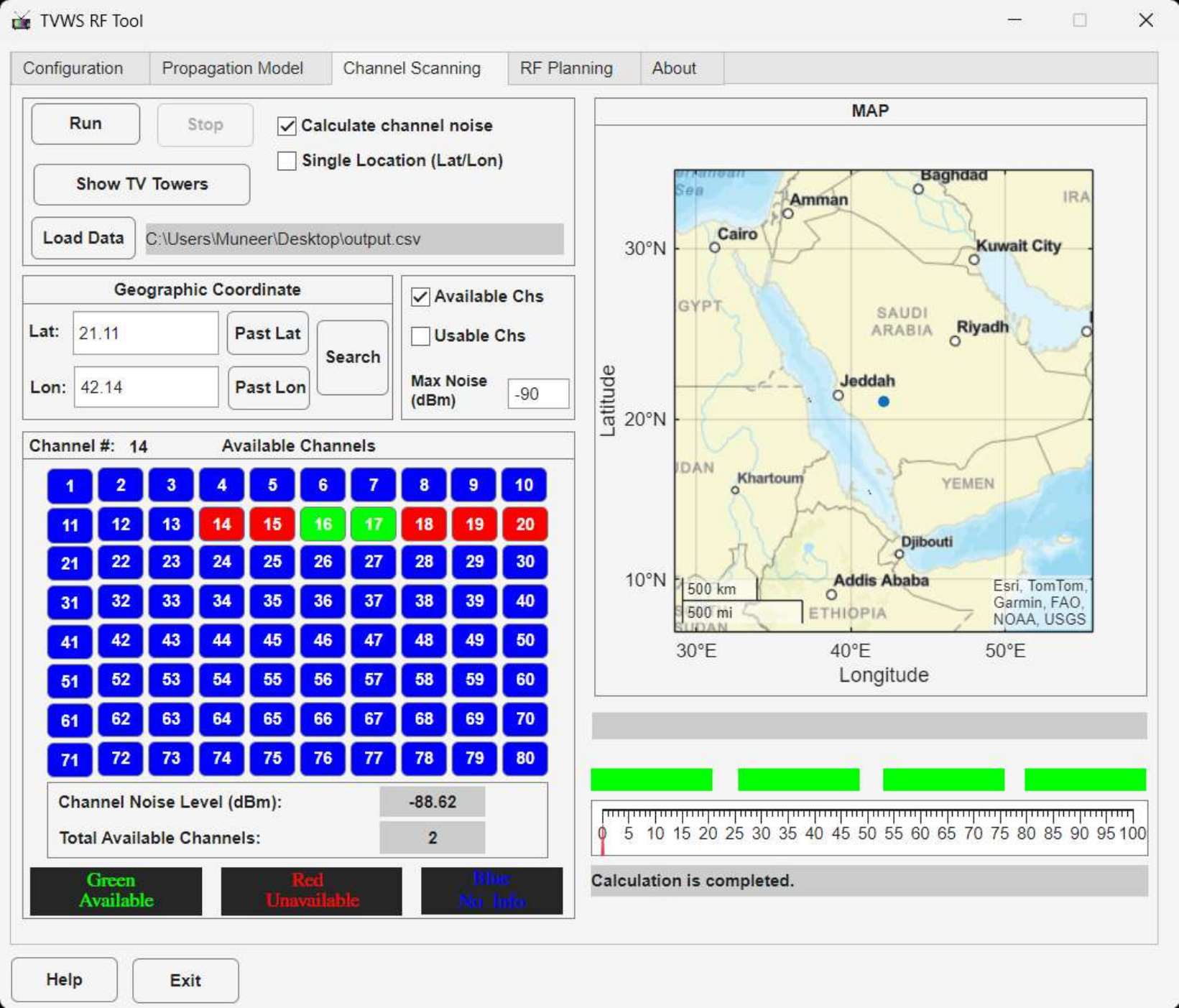}}
\caption{Sample of the displayed available and unavailable channels at a specific location.}
\label{fig:fig9}
\end{figure}

Then, to search the available/Usable channels at a specific location, the user can enter the coordinates in the designated "Lat" and "Lon" fields and then click the Search button. The tool will show the status of the scanned channels at this location using three different colours: Green for available and usable channels, Red for unavailable or unusable channels, and Blue to indicate channels with no status information, as shown in Fig. \ref{fig:fig9}. The channels with blue colour could represent the unscanned channels, or in case the entered location is outside the scanned region. In addition, the user can click on any channel number to show the noise level for that channel. Also, the total number of available and/or usable channels at this location will be displayed. 

\begin{figure}[htbp]
\centerline
{\includegraphics[width=\columnwidth]{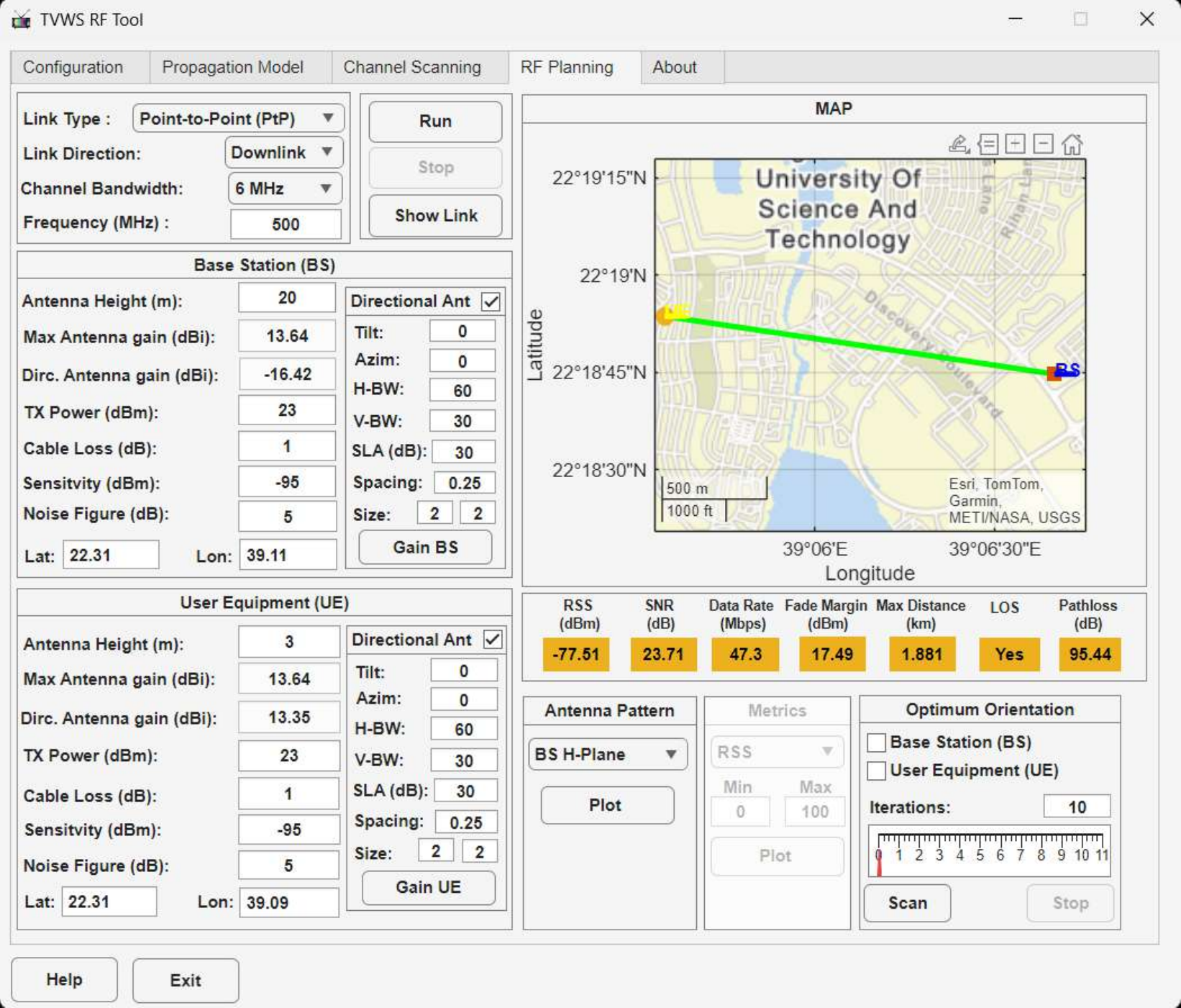}}
\caption{RF planning and optimization window.}
\label{fig:fig11}
\end{figure}

The RF planning and optimization part in the GUI tool, shown in Fig. \ref{fig:fig11}.  The user can choose either point-to-point (PtP) or point-to-multipoint (PtMP) scenarios over either downlink or uplink. The user can specify the TVWS carrier frequency and channel bandwidth. The characteristics of the BS and UE can be configured, including the coordinates, antenna height, transmit power, cable loss, sensitivity, and NF.  The default antenna for BS and UE is isotropic with a gain of 0 dBi.  However, the user can choose a directional antenna, i.e., uniform rectangular array (URA), by enabling the "Directional Ant" checkbox for BS and UE.  Then, the directional antenna parameters can be modified to get the desired radiation pattern. These parameters include the tilt angle in degrees (Tilt), azimuth angle in degrees (Azim),  the horizontal beam width (H-BW), the vertical beamwidth (V-BW),  maximum side-lobe level attenuation (SLA) in dB, elements spacing (Spacing), and array size of the elements (Size). For example, element spacing of 0.25 means that the elements are spaced by $\lambda/4$ where $\lambda$ is the wavelength. The radiation pattern of the BS and UE can be visualized by clicking the plot button in the "Antenna Pattern" panel. The user can select a vertical, horizontal, or 3D radiation pattern to be visualized. The user can click the Gain BS/UE button to calculate the maximum antenna gain. The directive antenna gain (Dirc. Antenna gain) represents the gain value in dB in the direction of the direct line connecting both BS and UE, which is obtained after running the simulation. Then, the user can click the Run button to evaluate the various communication metrics, which will be displayed in the designated fields. In addition, the map will show the location of BS/UE and the link status (LOS/NLOS), and the direction of both antennas. In the case of the PtMp scenario, the user can also plot the coverage color map for RSS, SNR, pathloss, FM, and data rate from the metrics panel. 
This tool also provides RF optimization techniques using Ray Tracing to find the best azimuth and elevation (tilt) orientation of the antenna for either BS or UE, and for both BS/UE antennas. This algorithm searches for the elevation/azimuth that provides the highest RSS at the receiver. To run the optimization technique, the user can click the Scan button in the optimum orientation panel. Then, the optimum elevation/azimuth will be displayed in the designated fields in the BS/UE antenna panels. 

\section{CONCLUSION}
In this work, we developed a generalized and flexible GUI tool to evaluate the TVWS spectrum by finding the status of TVWS channels and their noise levels at each geographic location within the analyzed area. Moreover, we develop an RF planning and optimization algorithm in this tool to evaluate the various communication metrics and find the optimal orientation for antennas. We have illustrated the methodology and criteria used to develop this tool in detail.  This tool can be used for any geographic region around the world by using the TV tower dataset and the realistic terrain profile of that region, with the flexibility to choose the desired spatial accuracy and resolution.  It allows the user to control the various system parameters, including the regulation rule parameters, the geographic area characteristics,  the channel range that needs to be scanned, the channel bandwidth, the reserved channels,  and others.  Moreover, it supports two widely used terrain-based radio propagation models, namely, the Longley-Rice, TIREM™, and Ray Tracing models.  This tool allows the user to export the output data as an external dataset file. Also, it can visualize the status of the TVWS channels (i.e., available, unavailable, usable, unusable, available \& usable),   their noise levels, and the total number of available/usable channels at any location entered by the user interactively.  The output data generated by the tool can be used to integrate with the cloud-based WSDB and to be visualised in other interactive maps using geospatial map tools. The RF planning feature provides the communication metrics between the TVWS BS and the UEs (e.g., WSDs), including received signal strength (RSS), pathloss, signal-to-noise ratio (SNR), data rate, fade margin, and line-of-sight (LOS) availability. Also, this tool has an RF optimization technique to find the best azimuth and elevation orientation for both the BS and the UE antennas using the Ray Tracing propagation model.

\bibliographystyle{IEEEtran}

\begin{thebibliography}{10}
\providecommand{\url}[1]{#1}
\csname url@samestyle\endcsname
\providecommand{\newblock}{\relax}
\providecommand{\bibinfo}[2]{#2}
\providecommand{\BIBentrySTDinterwordspacing}{\spaceskip=0pt\relax}
\providecommand{\BIBentryALTinterwordstretchfactor}{4}
\providecommand{\BIBentryALTinterwordspacing}{\spaceskip=\fontdimen2\font plus
\BIBentryALTinterwordstretchfactor\fontdimen3\font minus \fontdimen4\font\relax}
\providecommand{\BIBforeignlanguage}[2]{{%
\expandafter\ifx\csname l@#1\endcsname\relax
\typeout{** WARNING: IEEEtran.bst: No hyphenation pattern has been}%
\typeout{** loaded for the language `#1'. Using the pattern for}%
\typeout{** the default language instead.}%
\else
\language=\csname l@#1\endcsname
\fi
#2}}
\providecommand{\BIBdecl}{\relax}
\BIBdecl

\bibitem{r1}
M.~M. Al-Zubi and M.-S. Alouini, ``End-to-end modelling and simulation of nlos sub-6 ghz backhaul via diffraction for internet connectivity of rural areas,'' \emph{IEEE Open Journal of the Communications Society}, vol.~4, pp. 3102 -- 3114, 2023.

\bibitem{r17}
M.~Z. Islam, J.~F. O’Hara, D.~Shadoan, M.~Ibrahim, and S.~Ekin, ``Tv white space based wireless broadband internet connectivity: A case study with implementation details and performance analysis,'' \emph{IEEE Open Journal of the Communications Society}, vol.~2, pp. 2449--2462, 2021.

\bibitem{r18}
A.~Kumar, A.~Karandikar, G.~Naik, M.~Khaturia, S.~Saha, M.~Arora, and J.~Singh, ``Toward enabling broadband for a billion plus population with tv white spaces,'' \emph{IEEE Communications Magazine}, vol.~54, no.~7, pp. 28--34, 2016.

\bibitem{r19}
S.~W. Oh, Y.~Ma, M.-H. Tao, and E.~Peh, \emph{TV white space: The first step towards better utilization of frequency spectrum}.\hskip 1em plus 0.5em minus 0.4em\relax John Wiley \& Sons, 2016.

\bibitem{r2}
N.~A. Espinosa, N.~Arbel{\'a}ez, and G.~D. Castellanos, ``Recognition of available tv channels for tvws using sdr devices,'' in \emph{2017 IEEE Colombian Conference on Communications and Computing (COLCOM)}.\hskip 1em plus 0.5em minus 0.4em\relax IEEE, 2017, pp. 1--6.

\bibitem{r3}
F.~Awin, A.~Younan, D.~Corral-De-Witt, K.~Tepe, and E.~Abdel-Raheem, ``Real-time multi-channel tvws sensing prototype using software defined radio,'' in \emph{2018 IEEE International Symposium on Signal Processing and Information Technology (ISSPIT)}.\hskip 1em plus 0.5em minus 0.4em\relax IEEE, 2018, pp. 235--240.

\bibitem{r4}
O.~A. Oki, A.~Zulu, and M.~O. Adigun, ``Analysis of tv spectrum occupancy in kwadlangezwa township south africa,'' in \emph{2019 IEEE AFRICON}.\hskip 1em plus 0.5em minus 0.4em\relax IEEE, 2019, pp. 1--6.

\bibitem{r5}
M.~T. Masonta, D.~Johnson, and M.~Mzyece, ``The white space opportunity in southern africa: Measurements with meraka cognitive radio platform,'' in \emph{e-Infrastructure and e-Services for Developing Countries: Third International ICST Conference}.\hskip 1em plus 0.5em minus 0.4em\relax Springer, 2012, pp. 64--73.

\bibitem{r6}
N.~A. Espinosa, N.~Arbel{\'a}ez, and G.~D. Castellanos, ``Recognition of available tv channels for tvws using sdr devices,'' in \emph{2017 IEEE Colombian Conference on Communications and Computing (COLCOM)}.\hskip 1em plus 0.5em minus 0.4em\relax IEEE, 2017, pp. 1--6.

\bibitem{r7}
G.~Niringiye, I.~N. Oteyo, and T.~Bulega, ``Spectrum sensing for cognitive vhf land mobile radio communication networks using energy sensing techniques,'' in \emph{2021 IEEE AFRICON}.\hskip 1em plus 0.5em minus 0.4em\relax IEEE, 2021, pp. 1--6.

\bibitem{r23}
\BIBentryALTinterwordspacing
{Federal Communications Commission}. In the matter of unlicensed operation in the tv broadcast bands, additional spectrum for unlicensed devices below 900 mhz and in the 3 ghz band. (accessed Sep. 15, 2024). [Online]. Available: \url{https://www.fcc.gov/document/matter-unlicensed-operation-tv-broadcast-bands-additional}
\BIBentrySTDinterwordspacing

\bibitem{r8}
H.~Mohammed, T.~T. Terefe, and S.~Feisso, ``Coverage determination of incumbent system and available tv white space channels for secondary use in ethiopia,'' in \emph{Antenna Arrays-Applications to Modern Wireless and Space-Born Systems}.\hskip 1em plus 0.5em minus 0.4em\relax IntechOpen, 2021.

\bibitem{r9}
F.~Hessar and S.~Roy, ``Capacity considerations for secondary networks in tv white space,'' \emph{IEEE Transactions on Mobile Computing}, vol.~14, no.~9, pp. 1780--1793, 2014.

\bibitem{r10}
M.~Nekovee, ``Quantifying the availability of tv white spaces for cognitive radio operation in the uk,'' in \emph{2009 IEEE International Conference on Communications Workshops}.\hskip 1em plus 0.5em minus 0.4em\relax IEEE, 2009, pp. 1--5.

\bibitem{r11}
J.~Van De~Beek, J.~Riihijarvi, A.~Achtzehn, and P.~Mahonen, ``Tv white space in europe,'' \emph{IEEE Transactions on Mobile Computing}, vol.~11, no.~2, pp. 178--188, 2011.

\bibitem{r12}
J.~Van~de Beek, J.~Riihijarvi, A.~Achtzehn, and P.~Mahonen, ``Uhf white space in europe—a quantitative study into the potential of the 470--790 mhz band,'' in \emph{2011 IEEE International Symposium on Dynamic Spectrum Access Networks (DySPAN)}.\hskip 1em plus 0.5em minus 0.4em\relax IEEE, 2011, pp. 1--9.

\bibitem{r20}
G.~P. Villardi, H.~Harada, F.~Kojima, and H.~Yano, ``Primary contour prediction based on detailed topographic data and its impact on tv white space availability,'' \emph{IEEE Transactions on Antennas and Propagation}, vol.~64, no.~8, pp. 3619--3631, 2016.

\bibitem{r13}
G.~Naik, S.~Singhal, A.~Kumar, and A.~Karandikar, ``Quantitative assessment of tv white space in india,'' in \emph{2014 Twentieth National Conference on Communications (NCC)}.\hskip 1em plus 0.5em minus 0.4em\relax IEEE, 2014, pp. 1--6.

\bibitem{r14}
D.~Makris, G.~Gardikis, and A.~Kourtis, ``Quantifying tv white space capacity: A geolocation-based approach,'' \emph{IEEE Communications Magazine}, vol.~50, no.~9, p. 145, 2012.

\bibitem{r15}
H.~M. Hussien, K.~Katzis, L.~P. Mfupe, and T.~Ephrem, ``Calculation of tvws spectrum availability using geo-location white space spectrum database,'' in \emph{2021 IEEE AFRICON}.\hskip 1em plus 0.5em minus 0.4em\relax IEEE, 2021, pp. 1--6.

\bibitem{r16}
M.~Denkovska, P.~Latkoski, and L.~Gavrilovska, ``Geolocation database approach for secondary spectrum usage of tvws,'' in \emph{2011 19thTelecommunications Forum (TELFOR) Proceedings of Papers}.\hskip 1em plus 0.5em minus 0.4em\relax IEEE, 2011, pp. 369--372.

\bibitem{r24}
\BIBentryALTinterwordspacing
H.~I. Industries. Terrain integrated rough earth model (tirem). (accessed Sep. 16, 2024). [Online]. Available: \url{https://spectrum.hii-tsd.com/ModelStorefront/tirem?id=110}
\BIBentrySTDinterwordspacing

\bibitem{r21}
\BIBentryALTinterwordspacing
M.~M. Al-Zubi and M.-S. Alouini. {TVWS} channel scanner. (accessed Sep. 12, 2024). [Online]. Available: \url{https://github.com/muneer85/tvws-channels}
\BIBentrySTDinterwordspacing

\bibitem{r22}
------, ``Influence of digital terrain data accuracy on diffraction prediction in nlos wireless backhauls,'' \emph{IEEE Transactions on Vehicular Technology}, pp. 1--13, 2024.

\end{thebibliography}
\begin{bibliography}{refs}
\end{bibliography}

\begin{IEEEbiography}[{\includegraphics[width=1in,height=1.25in,clip,keepaspectratio]{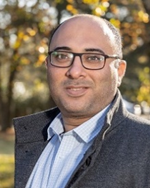}}]{MUNEER M. AL-ZUBI} ~received the Ph.D. degree in engineering from the University of Technology Sydney (UTS), Sydney, Australia, in 2020. He worked as a Research Associate with the Center of Excellence for Innovative Projects, Jordan University of Science and Technology (JUST), from 2020 to 2021. From 2021 to 2022, he was a postdoctoral researcher with the Department of Engineering, University of Luxembourg, Luxembourg, and also, he was a remote visiting scholar with the School of Electrical and Data Engineering, UTS. He is currently a Postdoctoral Researcher at the Communication Theory Lab (CTL), King Abdullah University of Science and Technology (KAUST), Saudi Arabia. His research interests lie in the areas of wireless communication, radio wave propagation, and molecular communication.
\end{IEEEbiography}

\begin{IEEEbiography}[{\includegraphics[width=1in,height=1.25in,clip,keepaspectratio]{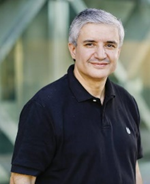}}]{MOHAMED-SLIM ALOUINI} ~{(S’94, M’98, SM’03, F’09)} ~received the Ph.D. degree in electrical engineering from the California Institute of Technology (Caltech), Pasadena, CA, USA, in 1998. He served as a faculty member at the University of Minnesota, Minneapolis, MN, USA, then at Texas A\&M University at Qatar, Education City, Doha, Qatar before joining King Abdullah University of Science and Technology (KAUST), Thuwal, Makkah Province, Saudi Arabia as a professor of electrical engineering in 2009. His current research interests include the modeling, design, and performance analysis of wireless communication systems.
\end{IEEEbiography}

\end{document}